\documentclass[
    ,final            
  ]
  {aipproc}

\layoutstyle{6x9}
\usepackage{sidecap}

\begin{document}

\title{Unitary and causal dynamics based \\ on the chiral Lagrangian}


\author{A.M. Gasparyan}{
  address={GSI Helmholtzzentrum f\"ur Schwerionenforschung GmbH, Planckstrasse 1, 64291 Darmstadt, Germany}
,altaddress={SSC RF ITEP, Bolshaya Cheremushkinskaya 25, 117218 Moscow,
 Russia}
}

\author{M.F.M. Lutz}{
  address={GSI Helmholtzzentrum f\"ur Schwerionenforschung GmbH, Planckstrasse 1, 64291 Darmstadt, Germany}
}

\begin{abstract}
Pion-nucleon scattering, pion photoproduction, and nucleon Compton scattering
are analyzed within a scheme based on the chiral Lagrangian.
Partial-wave amplitudes are obtained by an analytic extrapolation of subthreshold reaction amplitudes computed
in chiral perturbation theory, where the constraints set by  electromagnetic-gauge invariance, causality and
unitarity are used to stabilize the extrapolation. Experimental data  are reproduced up to
energies $\sqrt{s}\simeq 1300$ MeV in terms of the parameters relevant at order $Q^3$. A striking puzzle caused by
an old photon asymmetry measurement close to the pion production threshold is discussed.
\end{abstract}

\maketitle


\section{Introduction}
In recent years photon- and pion-nucleon interactions have been
successfully used as a quantitative challenge of chiral perturbation theory ($\chi $PT), which is a systematic tool
to learn about low-energy QCD dynamics \cite{Bernard:1995dp,Bernard:2007zu,Pascalutsa:2006up}.
The application of $\chi $PT is limited to the near threshold region. The pion-nucleon phase shifts have been analyzed
in great depth at subleading orders in the chiral expansion \cite{Bernard:1996gq,Fettes:1998ud,Fettes:2000xg}.
Pion photoproduction was studied in \cite{Bernard:1992nc,Bernard:1994gm,Bernard:1996ti,Fearing:2000uy}.
Compton scattering was considered in \cite{Bernard:1995dp,Beane:2004ra}.

The purpose of this talk is to report on a novel and unified description of  photon and pion scattering off the nucleon based on
the chiral Lagrangian \cite{Gasparyan:2010xz}. We aim at a description from threshold up to and beyond the isobar region in terms
of partial-wave amplitudes that are consistent with the constraints set by causality and unitarity.  Our analysis is based on the chiral
Lagrangian with pion and nucleon fields truncated at order $Q^3$. We do not consider an explicit isobar field in the chiral
Lagrangian. The physics of the isobar resonance enters our scheme by an infinite summation of higher order counter terms
in the chiral Lagrangian. The particular summation is performed in accordance with unitarity and causality.

The scheme is based on an analytic extrapolation of subthreshold scattering amplitudes that is controlled
by constraints set by electromagnetic-gauge invariance, causality and unitarity.
Unitarized scattering amplitudes are obtained which have left-hand cut structures in accordance with causality.
The latter are solutions of non-linear integral equations that are solved
by $N/D$ techniques. The integral equations are imposed on partial-wave amplitudes that are free of kinematical zeros and
singularities. An essential ingredient of the scheme is the analytic
continuation of the generalized potentials that determine the partial-wave amplitudes via the non-linear integral equation.
We discuss the analytic structure of the generalized potentials in detail and construct suitable conformal mappings in terms of
which the analytic continuation is performed systematically. Contributions from far distant left-hand cut structures are
represented by power series in the conformal variables.

The relevant counter terms of the Lagrangian are adjusted to the empirical data available for
photon and pion scattering off the nucleon. We focus on the s- and p-wave partial-wave amplitudes and
do not consider inelastic channels with two or more pions.
We recover the empirical s- and p-wave pion-nucleon
phase shifts up to about 1300 MeV quantitatively. The pion photoproduction process is analyzed in terms of its multipole
decomposition. Given the significant ambiguities in those multipoles we offer a more direct comparison of our results with
differential cross sections and polarization data. A quantitative reproduction of the data set up to energies of
about $\sqrt{s} \simeq 1300$ MeV is achieved.

\section{Analytic extrapolation of subthreshold scattering amplitudes}


The starting point of our method is the chiral Lagrangian involving pion,
nucleon and photon fields \cite{Fettes:1998ud,Bernard:2007zu}, for which we
collect the relevant terms
\begin{eqnarray}
\mathcal{L}_{int}&=&
-\frac{1}{4\,f^2}\,\bar{N}\,\gamma^{\mu}\,\big( \vec{\tau} \cdot \big(\vec{\pi}\times
(\partial_\mu\vec{\pi})\big)\big) \,N +
\frac{g_A}{2\,f} \,\bar{N}\,\gamma_5\,\gamma^{\mu} \,\big( \vec{\tau}\cdot (\partial_{\mu}\vec{\pi} )\big) \,N
\nonumber \\
&-&e\,\Big\{ \big(\vec{\pi}\times(\partial_{\mu}\vec{\pi}) \big)_3
+ \bar{N}\,\gamma_\mu\, \frac{1+\tau_3}{2} \,N
- \frac{g_A}{2\,f} \,\bar{N}\,\gamma_5\,\gamma_{\mu}\,\big(\vec\tau\times\vec{\pi}\big)_3\,N \Big\} \,A^\mu
\nonumber\\
&-&\frac{e}{4\,m_N}\,\bar{N}\,\sigma_{\mu\nu}\,\frac{\kappa_s+\kappa_v\,\tau_3}{2}\,N\,F^{\mu\nu}+
\frac{e^2}{32\pi^2 f}\,\epsilon^{\mu\nu\alpha\beta}\,\pi_3\,F_{\mu\nu}\,F_{\alpha\beta}
\nonumber\\
&-&\frac{2\,c_1}{f^2}\,m_\pi^2\, \bar{N}\,( \vec{\pi}\cdot\vec{\pi})\,N -
\frac{c_2}{2\,f^2\,m_N^2}\,\Big\{\bar{N}\,(
\partial_{\mu}\,\vec{\pi})\cdot (\partial_{\nu}\vec{\pi})\,(\partial^\mu \partial^\nu N )+\rm{h.c.}\Big\}
\nonumber \\
&+& \frac{c_3}{f^2}\,\bar{N} \,(\partial_{\mu}\,\vec{\pi} )
\cdot (\partial^{\mu}\vec{\pi})\,N
-\frac{c_4}{2\,f^2}\,\bar{N}\,\sigma^{\mu\nu}\,\big(\vec{\tau} \cdot \big((\partial_{\mu}\vec{\pi})\times
(\partial_{\nu}\vec{\pi})\big)\big)\,N
\nonumber\\
&-&i\,\frac{d_1+d_2}{f^2\,m_N}\,
\bar{N}\,\big(\vec\tau\cdot \big((\partial_\mu \vec \pi )\times
(\partial_\nu\partial_\mu \vec\pi) \big)\big) \, (\partial^\nu N) + \rm{h.c.}
\nonumber \\
&+&\frac{i\,d_3}{f^2\,m_N^3}\,
\bar{N}\,\big(\vec \tau \cdot \big( (\partial_\mu\vec\pi )\times
(\partial_\nu \partial_\lambda\vec\pi )\big)\big)\,
(\partial^\nu\partial^\mu\partial^\lambda  N)
+\mbox{h.c.}
\nonumber\\
&-&2\,i\,\frac{m_\pi^2\,d_5}{f^2\,m_N}\,\bar{N}\,\big(\vec{\tau} \cdot \big(\vec{\pi}\times
(\partial_\mu\vec{\pi}) \big)\big)\,( \partial ^\mu N) +\rm{h.c.}
\nonumber\\
&-&\frac{i \,e }{f\,m_N }\, \epsilon^{\mu\nu\alpha\beta}\,\bar{N}\,\big( d_8\,
(\partial_\alpha \,\pi_3)  +d_9\,\big(\vec\tau \cdot (\partial_\alpha \vec \pi)\big) \big)\, (\partial_\beta\,N)\, F_{\mu\nu}+\mbox{h.c.}
\nonumber \\
&+&i\,\frac{d_{14}-d_{15}}{2\,f^2\,m_N}\,
\bar{N}\,\sigma^{\mu\nu}\,\big((\partial_\nu\vec\pi )\cdot
(\partial_\mu\partial_\lambda\vec\pi ) \big)\,(\partial^\lambda N)
+\mbox{h.c.}
\nonumber \\
&-&\frac{m_\pi^2\,d_{18}}{f} \,\bar{N}\,\gamma_5\,\gamma^{\mu} \, \big( \vec{\tau}\cdot (\partial_{\mu}\vec{\pi})\big) \, N
\nonumber \\
&+&\frac{e \,(d_{22}-2\,d_{21})}{2\,f}\,\bar{N}\,\gamma_5\,\gamma^\mu\,
\big(\vec\tau\times\partial^\nu \,\vec \pi\big)_3 \,N\, F_{\mu\nu}
\nonumber \\
&+&\frac{e\, d_{20}}{2  \,f\,m_N^2}\,\bar{N}\,\gamma_5\,\gamma^\mu\,
\big(\vec\tau \times (\partial_\lambda \,\vec \pi)\big)_3\,
(\partial^\nu\partial^\lambda N)\, F_{\mu\nu}+\mbox{h.c.} \,.
\label{Lagrangian}
\end{eqnarray}
In a strict chiral expansion the order $Q^3$ results are composed from a tree-level part and a one-loop part.
Both parts are invariant under electromagnetic gauge transformations separately.
The tree-level pion-nucleon scattering amplitude receives contributions from the Weinberg-Tomozawa term,
the s- and u-channel nucleon  exchange processes and  the $Q^2$ and $Q^3$ counter terms characterized
by the parameters $c_1, ..., c_4$ and $d_1+d_2,d_3,d_5,d_{14}-d_{15},d_{18}$.
At tree-level the pion photoproduction amplitude is determined by the Kroll-Rudermann term, the
nucleon s- and u-channel exchange processes  and the t-channel pion exchange.
At chiral order $Q^3$ the counter terms $d_8,d_9,d_{18},2\,d_{21}-d_{22}$ contribute.
The tree-level amplitude for the proton Compton scattering is given
by nucleon s- and u-channel exchange contribution and pion t-channel exchange.

Our approach is based on  partial-wave dispersion relations, for which the unitarity and causality constraints can be
combined in an efficient manner. Using a partial-wave decomposition simplifies calculations because of
angular momentum and parity conservation. For a suitably chosen partial-wave amplitude with angular
momentum $J$, parity $P$ and channel quantum numbers $a,b$ we separate the right-hand cuts from the left-hand cuts
\begin{eqnarray}
 T_{ab}^{(JP)}(\sqrt{s}\,)=U_{ab}^{(JP)}(\sqrt{s}\,)+\int_{\mu_{\rm thr}}^{\infty}\frac{dw}{\pi}\frac{\sqrt{s}-\mu_M}{w-\mu_M}
\frac{\Delta T_{ab}^{(JP)}(w)}{w-\sqrt{s}-i\epsilon}\,,
\label{disrel}
\end{eqnarray}
where the generalized potential, $U_{ab}^{(JP)}(\sqrt{s}\,)$, contains left-hand cuts only, by definition.
The separation (\ref{disrel}) is gauge invariant. This follows since both contributions
in (\ref{disrel}) are strictly on-shell and characterized by distinct analytic properties.
The amplitude is considered as a function of $\sqrt{s}$ due to the MacDowell relations \cite{MacDowell:1959zza}.
A subtraction is made at $\sqrt{s}=\mu_M = m_N$ for the reasons to be discussed below.

The condition that the  scattering amplitude must be unitary allows one to calculate the discontinuity along
the right-hand  cut
\begin{eqnarray}
 \nonumber \Delta T_{ab}^{(JP)}(\sqrt{s})&=&\frac{1}{2\,i}\left(T_{ab}^{(JP)}(\sqrt{s}+i\epsilon)-
T_{ab}^{(JP)}(\sqrt{s}-i\epsilon)\right)\\
&=&\sum_{c,d}\,T_{ac}^{(JP)}(\sqrt{s}+i\epsilon)\,\rho^{(JP)}_{cd}(\sqrt{s}\,)\,T_{db}^{(JP)}(\sqrt{s}-i\epsilon),
\label{discontinuity}
\end{eqnarray}
where, $\rho^{(JP)}_{cd}(\sqrt{s}\,)$, is the phase-space matrix. The relation (\ref{disrel}) illustrates that the amplitude
possesses a unitarity cut along the positive real axis starting from the lowest $s$-channel threshold. In our case
the $\gamma \,N$ intermediate states induce a branch point at $\sqrt{s} = \mu_{\rm thr} = m_N$, which defines the
lowest s-channel unitarity threshold.

The structure of the left-hand cuts in $U_{ab}^{(JP)}(\sqrt{s}\,)$ can
be obtained by assuming the Mandelstam representation \cite{Mandelstam:1958xc,Ball:1961zza}
or examining the structure of Feynman diagrams in perturbation theory \cite{Nakanishi:1962}.
At leading order one may try to identify the generalized potential, $U_{ab}(\sqrt{s}\,)$, with a partial-wave projected
tree-level amplitude.  After all the tree-level expressions do not show any right-hand unitarity
cuts. However, this is an ill-defined strategy since it would lead to an unbounded
generalized potential for which the non-linear system (\ref{disrel}, \ref{discontinuity}) does not allow any solution.

The key observation is the fact that the solution of (\ref{disrel}, \ref{discontinuity})
requires the knowledge of the generalized potential for $\sqrt{s} > \mu_{\rm thr}$ only. Conformal mapping
techniques \cite{Frazer:1961zz} may be used to approximate the generalized potential in that domain efficiently based on the knowledge
of the generalized potential in a small subthreshold region around some expansion point $\mu_E$ only, where
it may be computed reliably in $\chi $PT. We establish a representation of the generalized potential of the form
\begin{eqnarray}
&& U(\sqrt{s}\,)=U_{\rm inside}(\sqrt{s}\,)+U_{\rm outside}(\sqrt{s}\,)\,,
\label{expansion} \\
&& U_{\rm outside}(\sqrt{s}\,)=\sum\limits_{k=0}^\infty {U_k}\,\big[\xi(\sqrt{s}\,)\big]^k\,, \qquad
U_k=\frac{d^k U_{\rm outside}(\xi^{-1}(\xi))}{k!\,d\xi^k}\Big|_{\xi=0}\,,
\nonumber
\end{eqnarray}
where we allow for an explicit treatment of left-hand cut structures that are inside a given domain $\Omega$.
The conformal mapping $\xi(\sqrt{s}\,)$ is defined unambiguously with
\begin{eqnarray}
\xi(\mu_E\,) =0\,,
\end{eqnarray}
and by the region $\Omega$, in which the expansion of $U_{\rm outside}(\sqrt{s}\,)$ is converging. The region $\Omega $ includes the real
interval $(\mu_E, \Lambda)$ where the parameter $\Lambda < \infty$ locates the opening of further inelastic channels not
considered explicitly. For
$ \sqrt{s}> \Lambda $ the outside part of the potential is set to a constant for convenience, where a suitable construction of
the conformal mapping  leads to a continuous behavior of $U_{\rm outside}(\sqrt{s}\,)$ and
$U'_{\rm outside}(\sqrt{s}\,)$ at $\sqrt{s} = \Lambda$. It is important to realize that the generalized potential is a smooth and
analytic function for $\mu_E <\sqrt{s} < \Lambda$.

For the elastic $\pi N \to \pi N$ potential the presence of baryon resonances
is encoded in the behavior of the potential at $\sqrt{s} < m_N - m_\pi $, a region not required to solve the non-linear system
(\ref{disrel}, \ref{discontinuity}). While at the matching scale $\sqrt{s} \simeq \mu_M$ we assume a perturbative behavior
of the generalized potential, this is no longer justified as the energy $\sqrt{s}$ is decreasing down to the region characterized
by the u-channel exchange of baryon resonances. Thus it is natural to insist on that the domain $\Omega$, excludes the u-channel
unitarity branch cuts. Such a construction lives in harmony with the crossing symmetry of the scattering amplitude.
Given the assumption the generalized potential cannot be evaluated by means of the representation (\ref{expansion}) for
$\sqrt{s} < m_N-m_\pi$.  The decomposition (\ref{expansion}) is faithful for energies $\sqrt{s} \in \Omega$ only. Nevertheless,
the full scattering amplitude can be reconstructed  at $\sqrt{s}\leq m_N-m_\pi$ from the knowledge of the generalized potential at
$\sqrt{s}> m_N+m_\pi$ by a crossing transformation of the solution to (\ref{disrel}, \ref{discontinuity}). Such a construction is
consistent with crossing symmetry if the solution to (\ref{disrel}, \ref{discontinuity}) and its crossing transformed form coincide in
the region $m_N-m_\pi < \sqrt{s} < m_N+m_\pi$.
This is the case approximatively, if the matching scale $\mu_M$ in  (\ref{disrel}) is identified properly.
For $m_N-m_\pi < \mu_M < m_N+m_\pi$ the scattering amplitude remains perturbative in the matching interval, at least sufficiently
close to $\sqrt{s} \sim \mu_M$. With our choice
\begin{eqnarray}
\mu_M = m_N \,,
\label{def-muM}
\end{eqnarray}
this is clearly the case (see also \cite{Lutz:2001yb}).

\begin{SCfigure}[][t]
\includegraphics*[width=11.5cm]{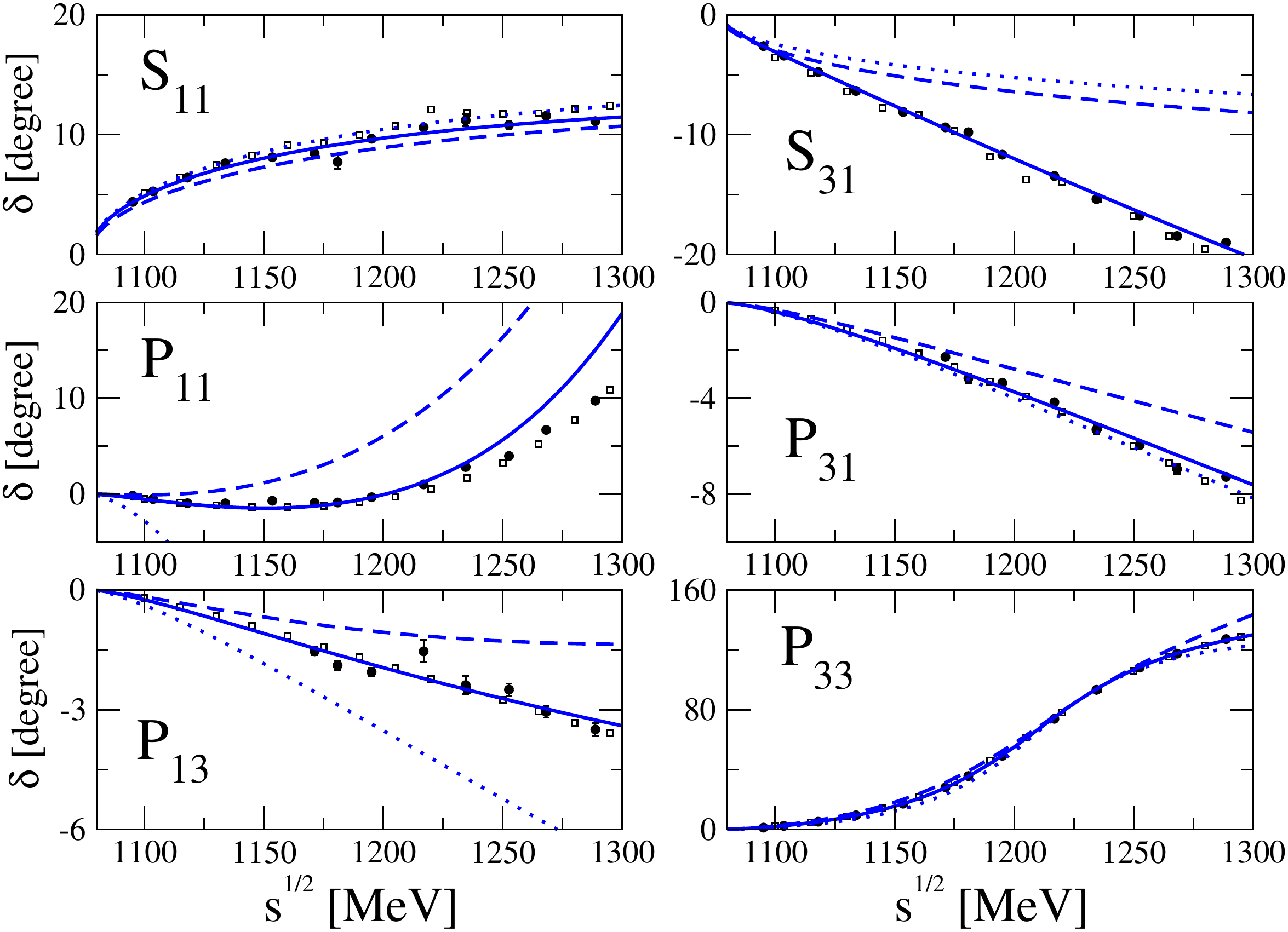}
\caption{Results of the fit for $\pi N$ $S$ and $P$-wave phase shifts.
The solid curves correspond to the full $Q^3$ results, the dashed curves
to $Q^2$ results, and the dotted curves to $Q^1$ calculation.
The data are from \cite{Koch:1985bn}(circles) and \cite{Arndt:2006bf}(squares).}
\label{fig:piN1}
\end{SCfigure}

\section{Results}

We performed numerical calculations for the three reactions $\pi N \to \pi N$, $\gamma N \to \pi N$ and
$\gamma N \to \gamma N$, where we assume the $\pi N$ channel in an s- or p-wave state \cite{Gasparyan:2010xz}.
The considered energy region from threshold up to $1.3$ GeV is motivated by the
expected reliability of the two-channel approximation \cite{Koch:1985bn,Arndt:2006bf}. The $\pi N$ sector is
developed independently from the $\gamma N$ channel
as it is treated to first order in the electric charge. After adjustment of the free parameters we obtain a good description
of the empirical phase shifts. There are four  $Q^2$  counter terms $c_{1-4}$. Further four
parameters $d_1+ d_2$, $d_{3,5}$ and $d_{14-15}$ are relevant at the one-loop level $Q^3$. As can be seen
from Fig.~\ref{fig:piN1} for all partial waves we obtain a convincing convergence pattern. For a more detailed discussion,
in particular the presence of CDD poles in the $P_{11}$ and $P_{33}$ waves we refer to
\cite{Gasparyan:2010xz}.

\begin{figure}[t]
 \includegraphics*[width=7.5cm]{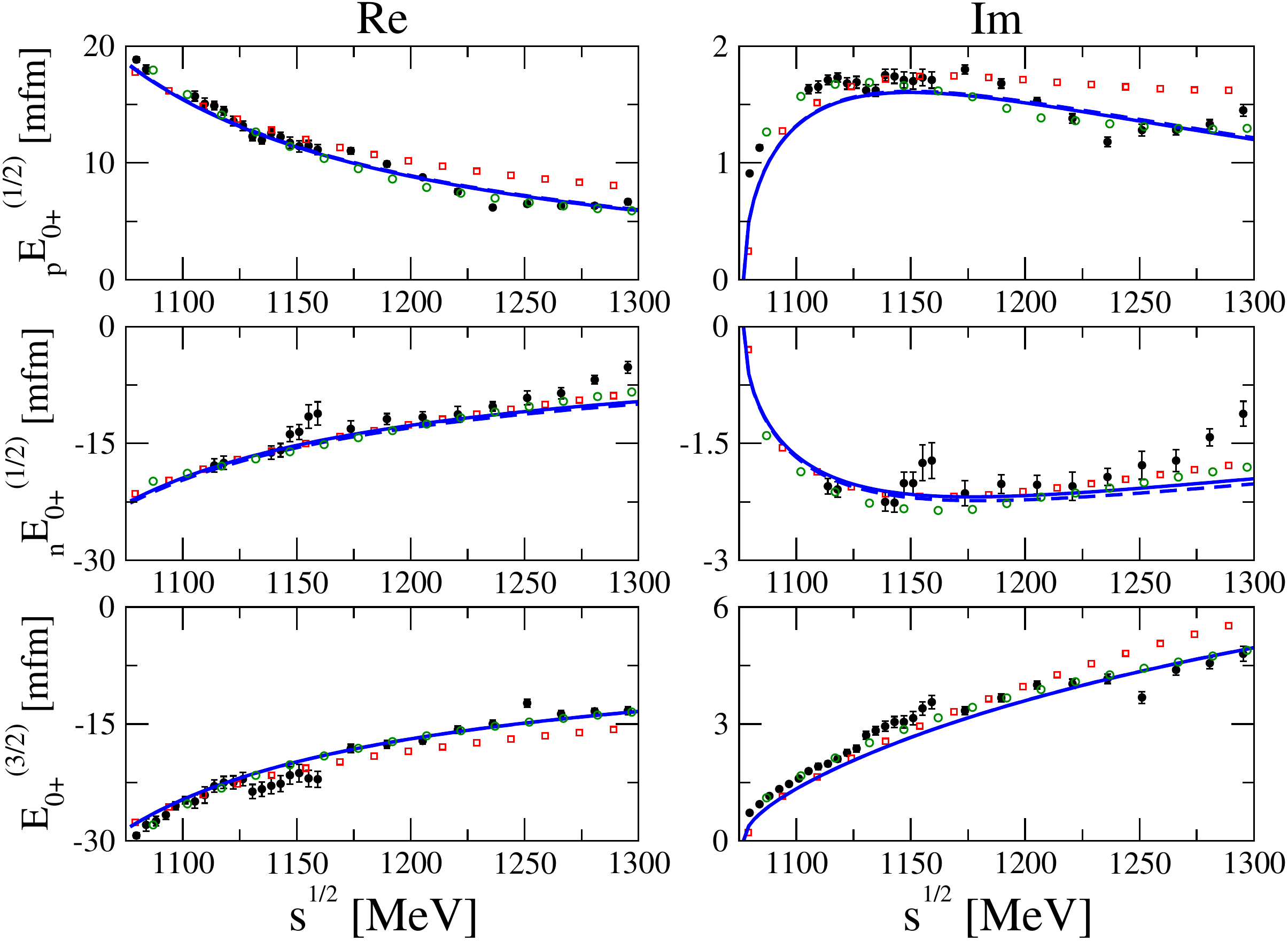}
   \includegraphics*[width=7.5cm]{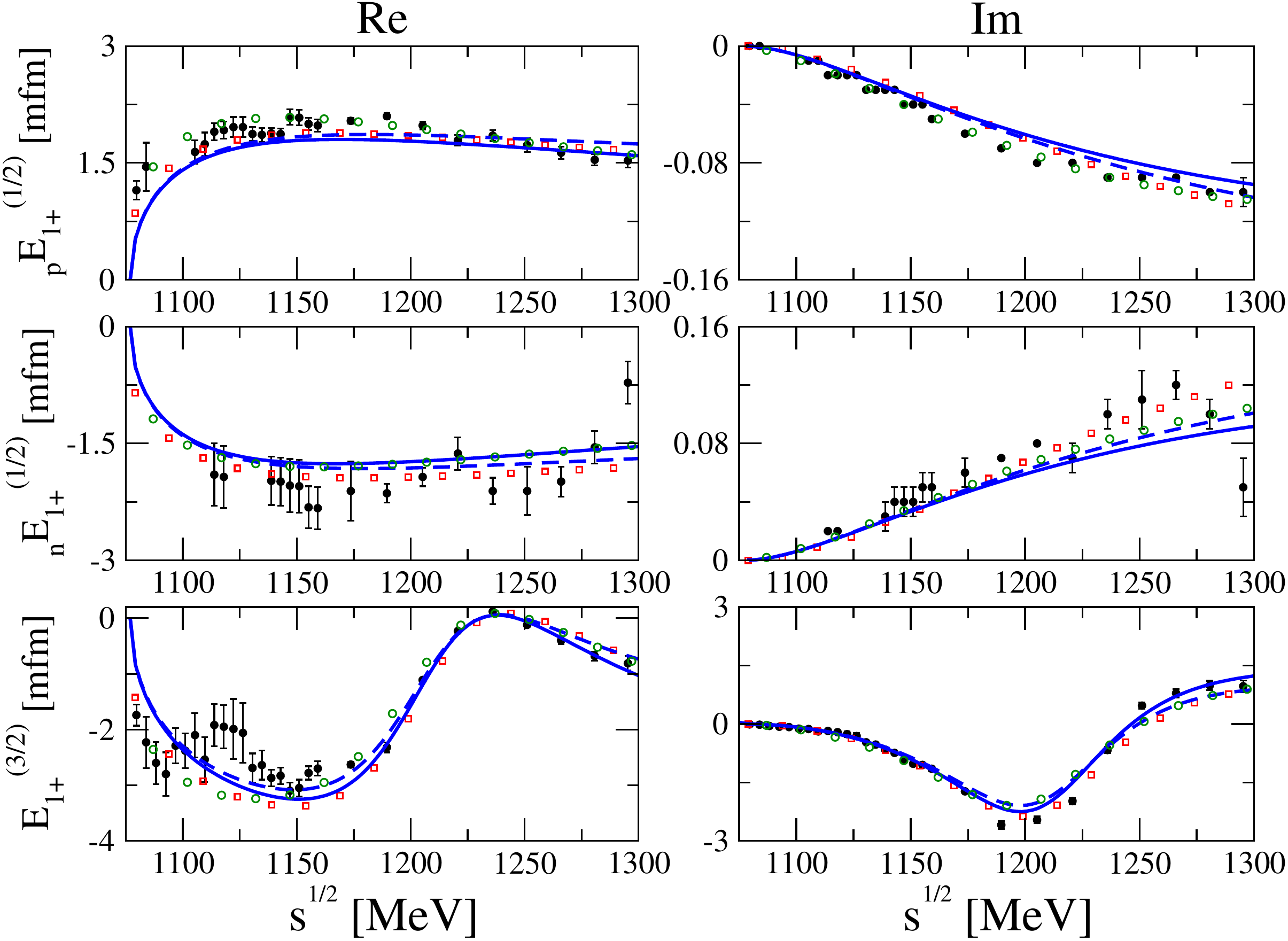}
\caption{Electric photoproduction multipoles $E_{0+}$ and $E_{1+}$. The data are from
\cite{Arndt:2002xv}(filled circles -- energy independent analyses, open circles
-- energy dependent analysis) and \cite{Drechsel:2007if}(squares). The solid line denotes the $Q^3$
calculation, the dashed line -- $Q^2$ calculation.}
\label{fig:gammaN1}
\end{figure}

We turn to pion photoproduction, which requires the determination of additional parameters. Besides further parameters
describing the coupling of the CDD poles to the $\gamma N$ channel there are additional counter terms of the chiral Lagrangian
to be considered. The relevant parameter set is obtained in part from the empirical s- and p-wave multipoles. There are
significant discrepancies among different energy-dependent analyses \cite{Arndt:2002xv,Drechsel:2007if}. Therefore we use the
energy independent partial-wave analysis from \cite{Arndt:2002xv}, which is less biased than the
energy dependent multipole analyses. Since the imaginary parts of the multipoles are given by Watson's
theorem \cite{Watson:1954uc}, their real parts are considered only. We find the $\pi N$ rescattering effects
important for the real parts of the s-wave multipoles. For the $p$-wave multipoles, which do not have a CDD pole
contribution, rescattering effects are mostly responsible for generating the correct imaginary part, but do not modify
the real parts much. The s-wave multipoles shown in Fig.~\ref{fig:gammaN1}
depend on the particular counter term combination
\begin{eqnarray}
\bar d_{20}+\frac{2\, \bar d_{21}-\bar d_{22}}{2}\,.
\label{s-wave-multipole}
\end{eqnarray}
Since none of the other multipoles depend on $\bar d_{20}$, the parameter combination (\ref{s-wave-multipole}) is determined
by the empirical s-wave multipoles. Given the spread in the different energy dependent  analyses a satisfactory description of the
s-wave multipoles is obtained.

\begin{figure}[t]
 \includegraphics*[width=7.5cm]{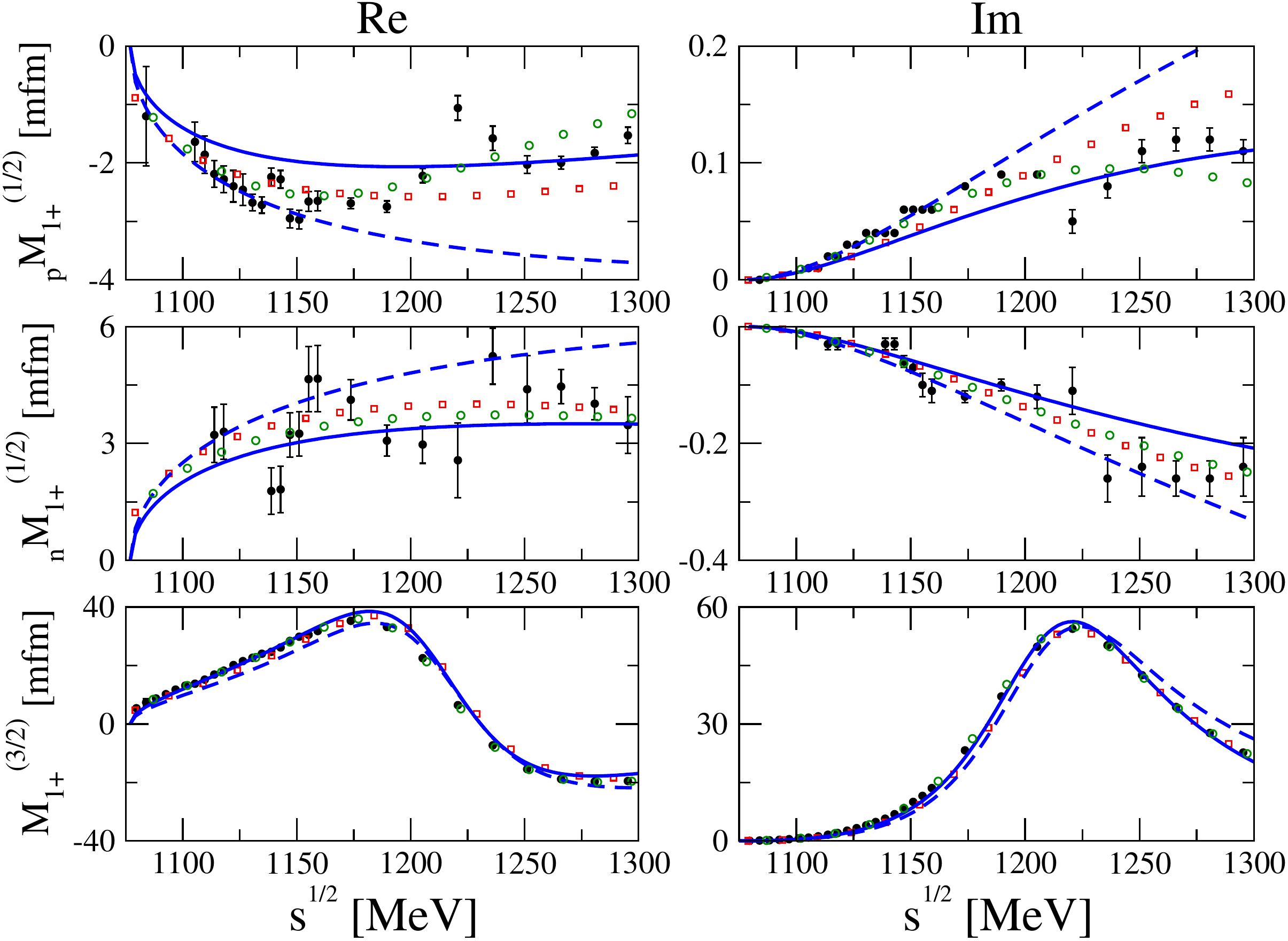}
  \includegraphics*[width=7.5cm]{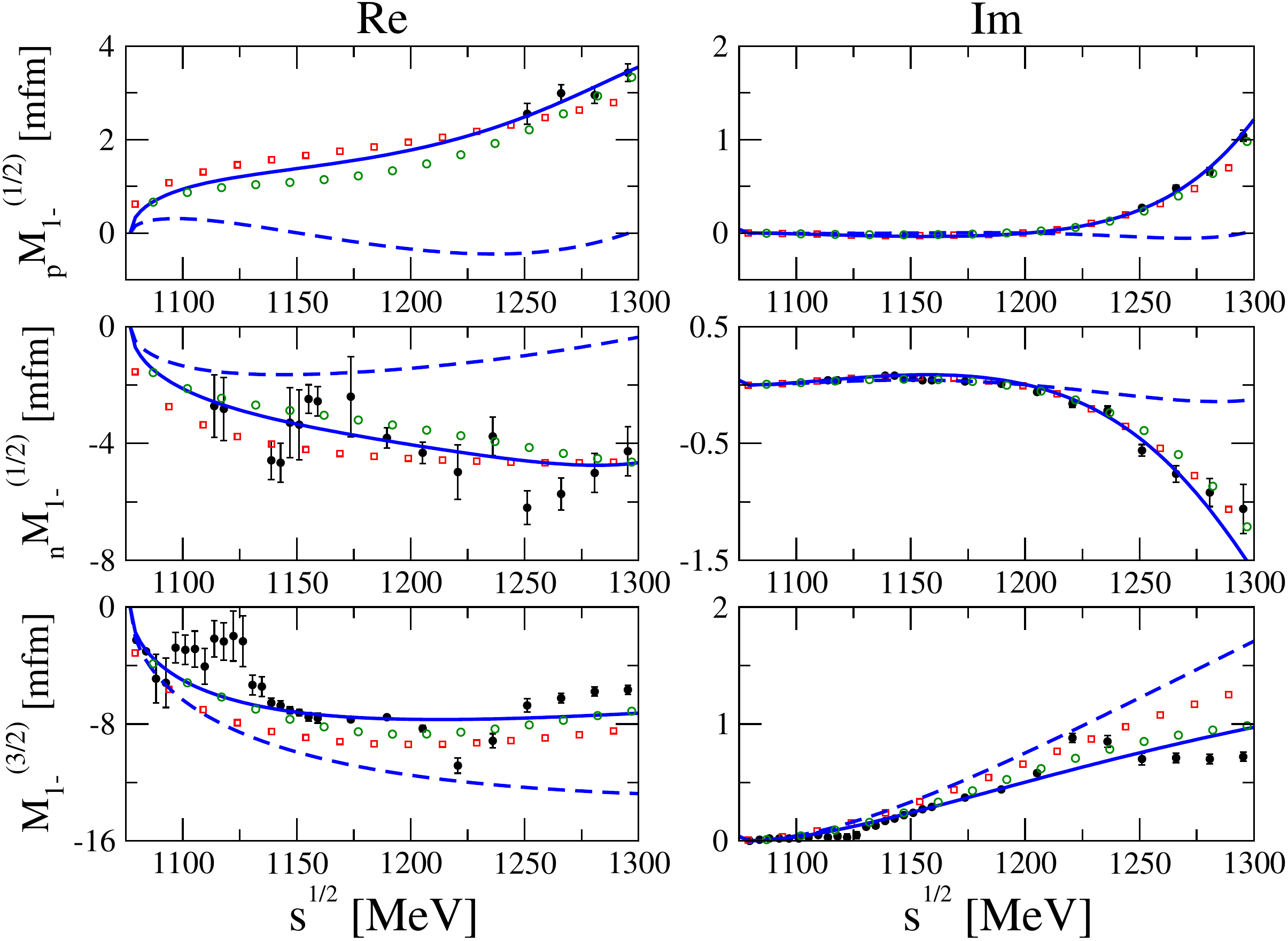}
\caption{Magnetic photoproduction multipoles $M_{1+}$ and $M_{1-}$. The data and line conventions are as in
Fig.~\ref{fig:gammaN1}.}
\label{fig:gammaN3}
\end{figure}

The magnetic multipole $M_{1+}^{(3/2)}$ multipole provides a large and dominant contribution to the cross sections
in the $\Delta (1232)$ resonance region \cite{Arndt:2002xv,Drechsel:2007if}. As a  consequence the empirical
error bars for this multipole are very small. The multipole depends on the particular parameter combination
\begin{eqnarray}
\bar d_8+\frac{2\, \bar d_{21}-\bar d_{22}}{2} \,,
\label{Delta-multipole}
\end{eqnarray}
which is relevant also for the electric multipole $E_{1+}^{(3/2)}$. The solid lines
in Fig.~\ref{fig:gammaN1} and Fig.~\ref{fig:gammaN3} show that a satisfactory description for both multipoles
is obtained.

There remain two parameter combinations, $\bar d_9$ and $2\, \bar d_{21}-\bar d_{22}$,
that cannot be determined unambiguously from the energy independent multipole analysis \cite{Arndt:2002xv}.
Incidentally, the energy dependent analyses \cite{Arndt:2002xv,Drechsel:2007if} differ most significantly in the remaining
five magnetic multipoles of Fig. ~\ref{fig:gammaN3}, which are sensitive to the latter parameters. The determination of  our preferred
values took into account additional constraints from photoproduction cross sections and Compton scattering data.
The resulting five magnetic multipoles are shown in Fig.~\ref{fig:gammaN3}.

\begin{figure}[t]
 \includegraphics*[width=10.5cm]{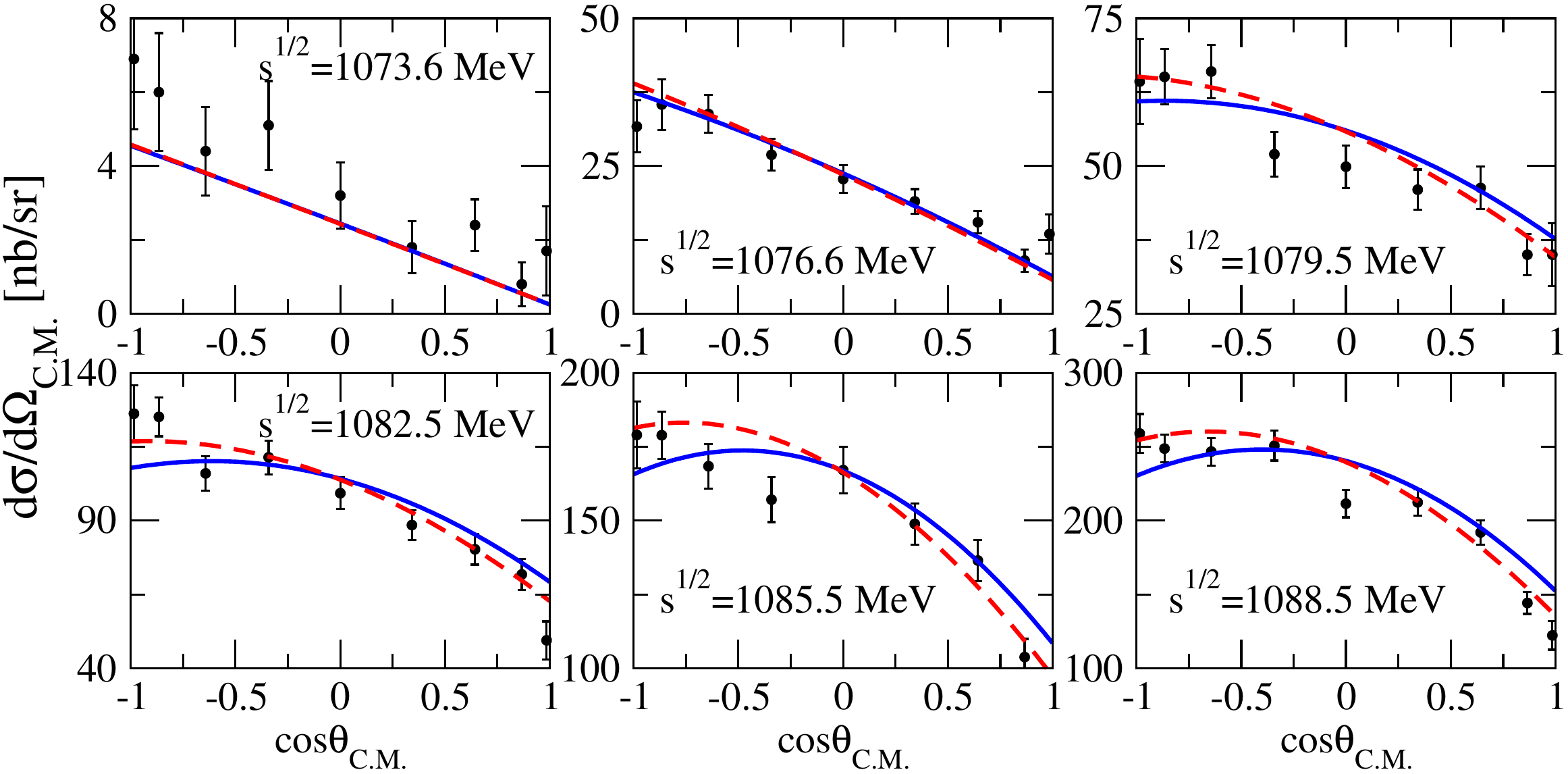}
\caption{Near threshold differential cross section for the
reaction $\gamma p\to\pi^0 p$ with data taken from \cite{Schmidt:2001,Schmidt:2001vg}.
Shown are results from  our coupled-channel theory including isospin breaking effects as are implied by the use of empirical pion
and nucleon masses. The solid lines correspond to our calculation with only $s$- and $p$-wave multipoles
included. The effect of higher partial waves is shown by the dashed lines. }
\label{fig:pi0threshold}
\end{figure}

Empirically it is well established that in neutral pion photoproduction
there is a strong cusp effect at the $\pi^+\,n$ threshold  \cite{Bergstrom:1996fq,Schmidt:2001vg}. To obtain accurate results we depart
from the isospin formulation and perform a coupled-channel computation in the particle basis using physical masses for the nucleons and
pions. Isospin breaking effects are not considered for the generalized potential being estimated to be of minor importance. No additional
parameters arise. Since the threshold region shows an intriguing interplay of the s- and p-wave multipoles
we confront our theory in Fig. \ref{fig:pi0threshold} with empirical differential cross section of the Mainz group directly. Given
the fact that we did not fit the parameters to those data an excellent description is achieved.

To further scrutinize the threshold region it is customary to introduce the following three
combinations of p-wave multipole amplitudes,
\begin{eqnarray}
&& \bar p_{\rm cm}\,\bar P_1=3\,E_{1+}+M_{1+}-M_{1-}\,,\qquad
\bar p_{\rm cm}\,\bar P_2=3\,E_{1+}-M_{1+}+M_{1-}\,,
\nonumber\\
&&\bar p_{\rm cm}\, \bar P_3 =2\,M_{1+}+M_{1-}\,,
\end{eqnarray}
which all vanish at the production threshold with $\bar p_{\rm cm}=0$. At leading orders in a chiral expansion
$\bar P_1$ and $\bar P_2$  do not depend on any of the $Q^3$ counter terms. The threshold behavior of $\bar P_1$ and
$\bar P_2$ is predicted in terms of the electromagnetic charge, the pion-nucleon coupling constant and the masses of the pions and
nucleons only \cite{Bernard:1994gm,Bernard:1995cj,Bernard:2001gz}. In contrast $\bar P_3$ receives a contribution
from the $Q^3$ counter term combination $\bar d_8 + \bar d_9$ and there is no parameter-free prediction
accurate to order $Q^3$. Since in our analysis all counter terms have been adjusted to photo-production data excluding
the near threshold region we can predict the threshold values of the three amplitude $\bar P_{1-3}$. We observe a striking disagreement
with the values obtained in \cite{Schmidt:2001vg}.

\begin{figure}[b]
 \includegraphics*[width=6.0cm]{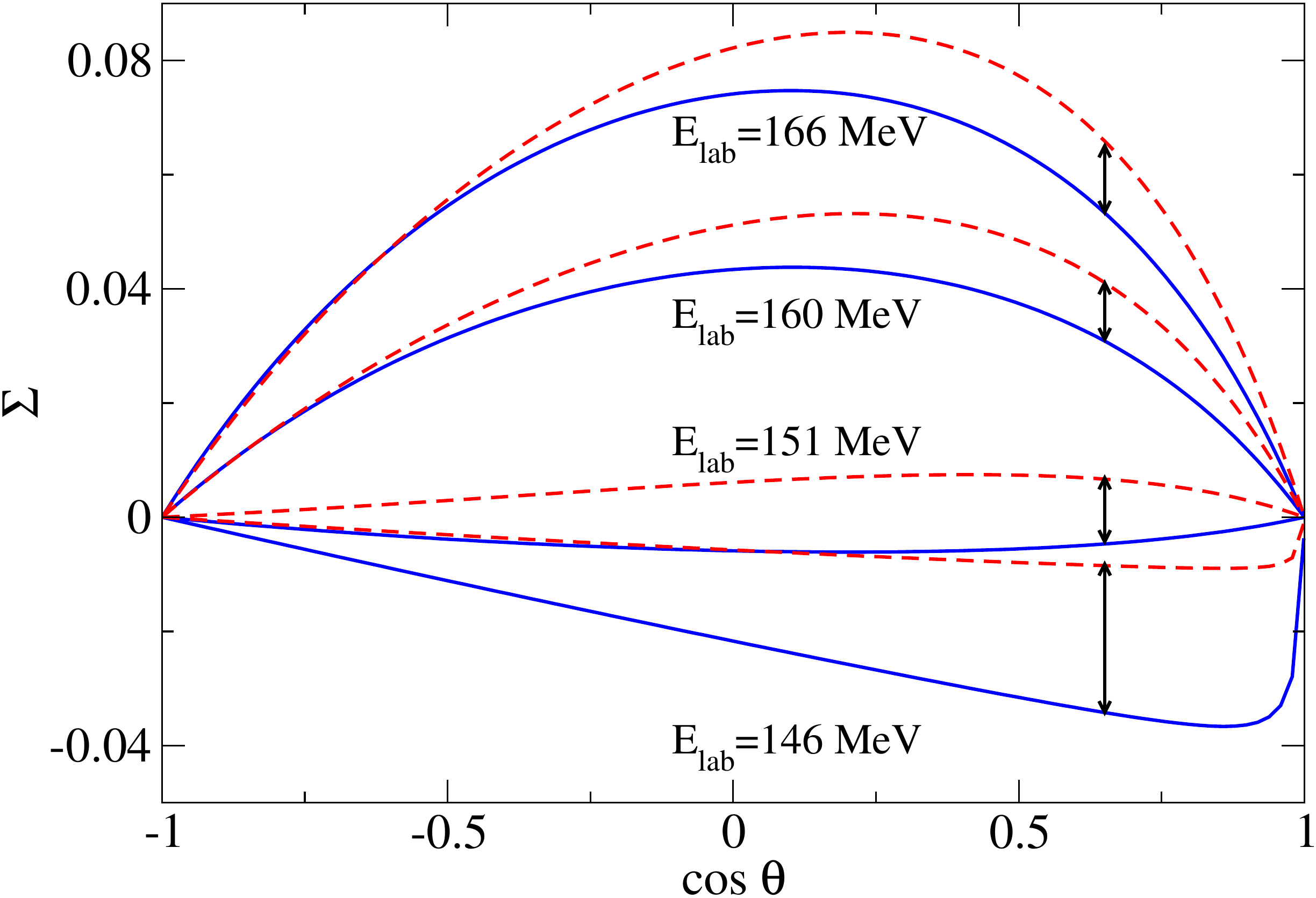}
\caption{Energy dependence of the photon asymmetry in
neutral pion photoproduction from the coupled-channel theory.
The solid lines correspond to our calculation with only $s$- and $p$-wave multipoles
included. The effect of higher partial waves is shown by the dashed lines.}
\label{fig:photon-asymmetry}
\end{figure}

From the near-threshold differential cross section two combinations of p-wave threshold parameters may be extracted. A complete
determination of all three threshold amplitudes requires additional information. For this purpose a measurement
of the near-threshold photon asymmetry suffices \cite{Schmidt:2001vg}. A direct comparison with the near-threshold photon
asymmetry measurement of the Mainz group \cite{Schmidt:2001vg} is not straightforward since an average  from threshold to
166 MeV laboratory photon energies was performed. In Fig. \ref{fig:photon-asymmetry} we show the result of our computation
at four different energies demonstrating a sign change in the asymmetry at energies below the mean value of 159.5 MeV in the Mainz
experiment. At the mean energy our photon asymmetry is about a factor four to five smaller than the averaged value
presented in \cite{Schmidt:2001vg}. A significant effect of higher partial wave contributions on the asymmetry is illustrated
by the dashed lines in Fig. \ref{fig:photon-asymmetry}. The possible importance of d-wave amplitudes in the photon asymmetry was pointed
out in \cite{FernandezRamirez:2009jb} recently. Within our scheme we have no freedom to significantly increase
the asymmetry without destroying the successful description of the photoproduction multipoles at higher energies. It is
interesting to recall that a negative photon asymmetry was obtained also in  \cite{Hanstein:1996bd,Kamalov:2001qg}
based on a dispersion-relation analysis of photo-production data. Given such a sign change an average over the photon
energy depends on the very details of the averaging procedure. We refer to the talk given by Michael Ostrick in this conference,
which reported on preliminary new data on the photon asymmetry. The consistency of the new data set with the old Mainz measurement
is being discussed.

\clearpage

\section{Summary}

We reviewed recent progress on a unified description of pion and photon-nucleon scattering up to and
beyond the isobar resonance region. A novel scheme that combines causality, unitarity and gauge invariance
and that is based on the chiral Lagrangian was discussed. The focus of the talk was the photo-production process
with a striking puzzle caused by an old beam asymmetry measurement at MAMI.




\bibliographystyle{aipproc}   

\bibliography{sample}

\end{document}